# Gate-tunable Intrinsic Anomalous Hall Effect in Epitaxial MnBi$_2$Te$_4$ Films

Shanshan Liu,[1,2#] Jiexiang Yu,[3#] Enze Zhang,[1] Zihan Li,[1,2] Qiang Sun,[4] Yong Zhang,[5] Lun Li,[5] Minhao Zhao,[1,2] Pengliang Leng,[1,2] Xiangyu Cao,[1,2] Jin Zou,[4,6] Xufeng Kou,[5] Jiadong Zang,[7] Faxian Xiu[1,2,8,9*]

[1]State Key Laboratory of Surface Physics and Department of Physics, Fudan University, Shanghai 200433, China

[2]Shanghai Qi Zhi Institute, 41th Floor, AI Tower, No. 701 Yunjin Road, Xuhui District, Shanghai 200232, China

[3]School of Physical Science and Technology, Soochow University, Suzhou 215006, China

[4]Materials Engineering, The University of Queensland, Brisbane QLD 4072, Australia

[5]School of Information Science and Technology, ShanghaiTech University, Shanghai 201210, China

[6]Centre for Microscopy and Microanalysis, The University of Queensland, Brisbane QLD 4072, Australia

[7]Department of Physics and Astronomy, University of New Hampshire, Durham, New Hampshire 03824, USA.

[8]Institute for Nanoelectronic Devices and Quantum Computing, Fudan University, Shanghai 200433, China

[9]Shanghai Research Center for Quantum Sciences, Shanghai 201315, China

[#] These authors contributed equally to this work

[*]Correspondence and requests for materials should be addressed to F. X. (Email: Faxian@fudan.edu.cn).

**Anomalous Hall effect (AHE) is an important transport signature revealing topological properties of magnetic materials and their spin textures. Recently, antiferromagnetic MnBi$_2$Te$_4$ has been demonstrated to be an intrinsic magnetic topological insulator that exhibits quantum AHE in exfoliated nanoflakes. However, its complicated AHE behaviors may offer an opportunity for the unexplored correlation between magnetism and band structure. Here, we show the Berry curvature dominated intrinsic AHE in wafer-scale MnBi$_2$Te$_4$ thin films. By utilizing a high-dielectric SrTiO$_3$ as the back-gate, we unveil an ambipolar conduction and electron-hole carrier ($n$-$p$) transition in ~7 septuple layer MnBi$_2$Te$_4$. A quadratic relation between the saturated AHE resistance and longitudinal resistance suggests its intrinsic AHE mechanism. For ~3 septuple layer MnBi$_2$Te$_4$, however, the AHE reverses its sign from pristine negative to positive under the electric-gating. The first-principles calculations demonstrate that such behavior is due to the competing Berry curvature between polarized spin-minority-dominated surface states and spin-majority-dominated inner-bands. Our results shed light on the physical mechanism of the gate-tunable intrinsic AHE in MnBi$_2$Te$_4$ thin films and provide a feasible approach to engineering its AHE.**

# Introduction

The study of anomalous Hall effect (AHE) is fundamental yet critical for understanding electron properties and magnetic couplings in magnetic materials[1,2]. In contrast to the normal Hall effect induced by Lorentz force, the AHE in magnetic compounds, especially with the ferromagnetism, correlates with the crystal structure and spin-orbit coupling[1,3,4]. There are three established microscopic mechanisms. Momentum-space Berry curvature contributes to the intrinsic AHE that depends on band structures[5]. Side jump and skew-scattering, resulted from opposite electron deflection and asymmetric electron scattering when approaching impurities[6–8], constitute the extrinsic effects of AHE. Through tracking the AHE evolution, the interaction between carriers and spins was discovered in traditional diluted magnetic semiconductors (In, Mn)As[9]. Magnetic exchange interaction was also found to switch from the Ruderman-Kittel-Kasuya-Yoshida (RKKY) to the Van-Vleck effect owing to the enhanced direct magnetic coupling of local electrons in heavily-doped ferromagnetic films[10]. Recently, the AHE has been extended to the antiferromagnetic compounds with unique magnetic textures or band structures, for example, the large AHE in the noncollinear $Mn_3Sn$ antiferromagnet[11]. In topological semimetal antiferromagnetic RPtBi (R is a rare earth element), the AHE is originated from the large Berry curvature introduced by Weyl points and magnetic textures[12,13].

When combining topological insulators with magnetism, a new topological phase, magnetic topological insulators can be developed which hold a surface exchange gap induced by a long-range ferromagnetic order. Many exotic phenomena were predicted in magnetic topological insulators, for instance, the quantum anomalous Hall effect and magnetic monopoles[14–16]. Through doping magnetic atoms into topological insulators, quantum anomalous Hall effect and topological Hall effect were detected[17–21]. Also, by employing magnetic insulators as the substrate, topological insulators with perpendicular ferromagnetism can be constructed due to the interfacial proximity effect [22–24]. However, disorders induced by the doping or lattice mismatch are almost inevitable, and films with high uniformity and stable magnetism are highly pursued[20,25,26]. Meanwhile, searching for intrinsic magnetic topological insulators with pristine magnetism and topological band structure becomes indispensable. To date, $MnBi_2Te_4$ has been experimentally confirmed to be an intrinsic magnetic topological insulator[27–29]. Due to the A-type antiferromagnetic order, versatile quantum properties in odd/even layers have been detected[30–35]. However, these findings in $MnBi_2Te_4$ flakes are susceptible to other extrinsic factors, like sample quality, suggesting that disorders and magnetization can significantly tune the electronic properties. Moreover, $MnBi_2Te_4$ in bulk crystals and exfoliated flakes were extensively studied with rare films involved which show complex AHE behaviors[36–38]. How the band structure correlates with the AHE and what kind of magnetism coupling exist, RKKY or Van-Vleck effect, pose open questions in $MnBi_2Te_4$.

Here, we report the gate-tunable Berry-curvature-dominated anomalous Hall effect in $MnBi_2Te_4$ films. Wafer-scale high crystalline $MnBi_2Te_4$ thin films have been directly deposited on (001) $Al_2O_3$ and (111) $SrTiO_3$ substrates. The antiferromagnetism, perpendicular magnetic anisotropy, and anomalous Hall insulator state of $MnBi_2Te_4$ thin films are demonstrated. By controlling the gate voltage, we have witnessed the electron-hole carrier (*n-p*) transition in the ~7 septuple layer (SL) sample and the reversed AHE sign in the ~3 SL sample. The intrinsic AHE is confirmed to be irrelevant to the carrier type or density. Band structure and Berry curvature calculations find that as the Fermi energy increases, the main conduction band switches from spin-minority to spin-majority bands, and the anomalous Hall conductivity changes from positive to negative. Our first-principles calculations explain the experimental finding and map the gated Fermi energy within the band structure.

## Results

Layered tetradymite $MnBi_2Te_4$ is a trigonal structure with $Te^{2-}$-$Bi^{3+}$-$Te^{2-}$-$Mn^{2+}$-$Te^{2-}$-$Bi^{3+}$-$Te^{2-}$ septuple layer like MnTe bilayer intercalated into $Bi_2Te_3$ quintuple layer, as schematically depicted in Fig. 1a. The A-type antiferromagnetic (AF) order on Mn can be described as intralayer ferromagnetic (FM) and interlayer AF coupling, as marked by green arrows. We successfully synthesized 2-inch high-quality $MnBi_2Te_4$ films on (001) $Al_2O_3$ and (111) $SrTiO_3$ substrates using molecular beam epitaxy by co-evaporating three elements of Mn, Bi, and Te (Fig. 1b inset) and by adjusting the element flux ratio and substrate temperature. The oscillation of the reflection-high-energy-electron-diffraction (RHEED) intensity during the growth of a 14 SL film on $Al_2O_3$ is shown in Fig. 1b, with a composition of $MnBi_{2.03}Te_{3.94}$ (Fig. S2a). The layer-by-layer growth mode is verified via periodic oscillations with a growth rate of 310 s/layer. The upper right inset of Fig. 1b shows a streaky RHEED diffraction pattern, suggesting a smooth surface. In Fig. 1c, the layered structure is further characterized by cross-section high-resolution transmission electron microscopy (HRTEM) and the layer distance is about 1.39 nm. The corresponding selected-area-electron-diffraction (SAED) results further identify the single-crystallinity (Fig. 1d). The epitaxial orientation of {003} is also confirmed by X-ray diffraction (XRD), as shown in Fig. 1e. Note that due to the thickness of 14 SL, signals from $Al_2O_3$ can also be captured in XRD and SAED measurements. The lattice constants of the thin film are *a*=4.38 Å and *c*=40.71 Å, consistent with the bulk values[39]. The detailed structural analysis is provided in Supplementary Section 1. We further conducted the magnetization measurements of zero-field-cooled (*ZFC*) and field-cooled (*FC*) magnetization and field-dependent magnetization to explore its magnetic properties. As illustrated in Fig. 1f, our 34 SL $MnBi_2Te_4$ film possesses the AF characteristic with a Néel temperature ($T_N$) of ~25.2 K, close to the bulk value[40,41]. Multiple steps in *M-H* hysteresis further confirm its antiferromagnetic property (Fig. 1f inset). As the magnetic field increases above 5 T, spins of Mn are aligned along the field direction and the spin-polarized state can be generated. Employing the experimental lattice constants, the density functional theory (DFT) calculations for $MnBi_2Te_4$ bulk show that the A-type AF state is 1.5 meV per Mn lower in energy than the FM state, confirming the experimental observation.

To investigate transport properties, we carried out magneto-transport measurements on $MnBi_2Te_4$/$Al_2O_3$ films to measure four-terminal longitudinal resistance ($R_{xx}$) and anomalous Hall resistance ($R_{xy}^A$). We subtracted the linear component of the normal Hall resistance ($R_{xy}^H$). Figure 2a shows the $R_{xx}$ of a 9 SL film. $R_{xx}$ rises with temperature dropping so it behaves as an insulator. A phase transition temperature appears at 19.8 K and corresponds to its Néel temperature $T_N$. Due to the enlarged thermal fluctuation effect in thin films, $T_N$ is smaller than the bulk value[40,41]. The activation energy $E_a$ is about 9.6 meV by fitting $R_{xx}$ at high temperatures to $R_{xx} \sim \exp(E_a/k_B T)$ where $k_B$ is the Boltzmann constant. The band gap ($E_{gap}$) is roughly estimated to be larger than 19.2 meV. Angle-dependent AHE under 1.5 K is shown in Fig. 2b, with the angle ($\theta$) defined as the angle between the magnetic field and the normal orientation of the sample surface. Perpendicular magnetic anisotropy can be validated as the saturation field ($H_s$) rises monotonously with the increasing angle. It was later confirmed by the DFT calculations that the perpendicular geometry of the A-type AF order is 0.8 meV per Mn lower in energy than the in-plane spin directions. Because of the A-type AF order with weak interlayer coupling, Mn from different layers can flip their spins independently so the multiple magnetic intermediate spin-flop states with kinks are identified in the $R_{xy}^A$ curve. In the temperature-dependent AHE measurement with $\theta=0°$, both $R_{xy}^A$ and coercive field ($H_c$) decrease monotonously as the temperature rises. The saturated anomalous Hall resistance ($R_{xy}^{AS}$) is about 6.72 kΩ (1.5 K).

The thickness-dependent AHE is obtained for 7 SL to 14 SL $MnBi_2Te_4/Al_2O_3$, as shown in Fig. 2d. $R_{xy}^A$ as well as the hysteresis window decreases monotonously as the thickness increases. It is attributed to the larger magnetic anisotropy in thinner samples which have been reported in two-dimensional ferromagnetic materials[42,43]. To make this clear, the extracted temperature-dependent coercive field ($H_c$) is drawn in Fig. 2d inset. As the film becomes thicker, more spin-flop states (in some cases, degenerate) can be generated, and the differences between each adjacent spin-flop state become smaller (Fig. 2d). It is reported that the quantum anomalous Hall effect with nearly quantized $R_{xy}^A$ and quenched $R_{xx}$ can be observed in low-disorder co-doped topological insulators[44] (named as quantum anomalous Hall insulator). At the same time, the decreased $R_{xy}^A$ and increased $R_{xx}$ can also appear due to the increased magnetic disorder (denoted as anomalous Hall insulator). Such behaviors can be analyzed via the evolution of $R_{xx}$. Zero-field $R_{xx}$ ($R_{xx}^0$) and $R_{xx}$ at $H_c$ ($R_{xx}^{H_c}$) of different $MnBi_2Te_4$ films from 7 SL to 14 SL exhibit a positive trend (Fig. 2e) and indicates that the $MnBi_2Te_4$ films share the anomalous Hall insulator state with strong local magnetic disorder although the high crystallinity has been proved by structural characterizations.

We further investigated the AHE with the modulation of the Fermi-level to connect it with the band structure. A high dielectric $SrTiO_3$ is served as the back-gate and two $MnBi_2Te_4$ samples were synthesized on 0.25 mm-thick (111) $SrTiO_3$ substrates and fabricated into a six Hall-bar structure (Fig. 3c). The thickness of the two samples is 5 nm and 10 nm, estimated to be ~3 SL and ~7 SL, respectively. Both the pristine samples show the electron-dominated semiconducting properties (*n*-type, see Figs. S5 and S8). Let's first analyze the ~7 SL film. Figure 3a shows the representative Hall resistance $R_{xy}$ curves at different gate voltages ($V_g$), and the corresponding $R_{xy}^A$ results (Fig. 3b) are obtained by subtracting the linear component from the normal Hall effect. In the positive $V_g$, the negative slopes of $R_{xy}$ curves indicate the *n*-type conduction but the negative $V_g$ flips the sign of the slope, indicating hole-carrier (*p*- type) conduction. This electron-hole (*n-p*) transition suggests that the Fermi-level is moved from the conduction band to the valence band as $V_g$ decreases (Fig. 3d inset). The normal Hall resistance $R_{xy}^H = R_H B$ is proportional to the magnetic field $B$, so the sheet carrier density at each gate voltage can be extracted by $1/(R_H \cdot e)$ where $e$ is the elementary charge, as shown in Fig. 3d. With $V_g$ decreasing, the electron density shows a dramatically declining behavior while the hole density does not change much. Secondly, the hole density is larger than the electron density near the *n-p* transition. These findings indicate a narrow conduction band and a broad valence band, as schematically shown in Fig. 3d inset. To avoid the large errors of the estimated carrier density because of the complex Hall slope, we do not extract the carrier density near the critical *n-p* transition regime. The critical *n-p* transition regime is also verified by the longitudinal resistance $R_{xx}^0$ processed by the symmetric methods[30], as shown in Fig. 3e. $R_{xx}^0$ has an ambipolar behavior with a maximum of 13.86 kΩ at $V_g = 10$ V, and the asymmetric feature demonstrates the different properties between electron and hole carriers. We note that the carrier conduction switches from electrons to holes and the carrier density changes (Fig. 3d). The negative sign of $R_{xy}^A$ is preserved in the whole measurement. The small kink at about ±3 T corresponds to the magnetic transition. The $R_{xy}^{AS}$ also shows an ambipolar feature with a maximum of 844 Ω at $V_g = 20$ V. The relation between $R_{xy}^{AS}$ and $R_{xx}^0$ in both electron and hole regions follow the quadratic scaling of $R_{xy}^{AS} \propto (R_{xx}^0)^2$ (Fig. 3f). It thus gives a strong support for the intrinsic AHE dominated by Berry curvature in MnBi$_2$Te$_4$ films.

Distinct from the *n-p* transition in ~7 SL device, $R_{xy}$ of ~3 SL film behaves *n*-type conduction in the whole $V_g$ range of ±160 V, and the field-dependent $R_{xy}$ and the corresponding $R_{xy}^A$ curves at various gate voltages are illustrated in Figs. 4a and 4b, respectively. However, in contrast to the ~7 SL sample, $R_{xy}^{AS}$ in ~3 SL changes its sign at positive $V_g$ where $R_{xy}^{AS}$ is positive/negative under the positive/negative saturated magnetic field (Fig. 4b). We call respectively positive- and negative-sign AHE when $R_{xy}^{AS}$ has the same and opposite sign with the saturated magnetic field. With various $V_g$, the saturated $R_{xy}^{AS}$ and the corresponding $R_{xx}^0$ are shown in Fig. 4c. The AHE polarity switches its sign at $V_g =30$ V. The magnitudes of positive-sign AHE are much smaller than those of negative-sign AHE. $R_{xx}^0$ increases monotonously as $V_g$ scans from positive to negative. It indicates that the carrier density decreases as $V_g$ reduces, consistent with the evolution of $n_s$ extracted from the linear Hall slope (Fig. S8d). The quadratic relations $R_{xy}^{AS} \propto (R_{xx}^0)^2$ between $R_{xy}^{AS}$ and $R_{xx}^0$ are also verified for both positive- and negative-sign AHE, as shown in Fig. 4d, confirming the Berry curvature dominated intrinsic AHE. To check the magnetic tunability, we have exploited the temperature-dependent $R_{xy}$ under different gating voltages. We found that MnBi$_2$Te$_4$ films contain pristine and stable magnetism under different $V_g$, with the $T_N$ nearly unchanged irrespective of the carrier type and density (Supplementary Figs. S7, S10).

Due to the antiferromagnetic order, the even layer MnBi$_2$Te$_4$ always brings about zero AHE. Thus, by using first-principles calculations with the assistance of the maximally-localized Wannier functions method, the band structures and their corresponding intrinsic anomalous Hall conductivity $\sigma_{xy}^A$ were calculated for 3, 5, and 7 SL MnBi$_2$Te$_4$. Under the perpendicular antiferromagnetic spin order where the spins at the first and the last layer are set to be the up direction, the band structures are shown in Fig. 5a-c. For all three films, a bandgap of about 0.05 eV appears at the Fermi energy with the insulating nature, consistent with the activated energy $E_a$. The corresponding anomalous Hall conductivities with various Fermi energies are summarized in Fig. 5d. Inside the bandgap, the non-zero platform of $\sigma_{xy}^A$ appears for all three films. The magnitude of the platform is exactly $e^2/h$, indicating the quantum anomalous Hall effect. Above the gap, $\sigma_{xy}^A$ decays significantly as the Fermi energy increases. Valence bands are narrower than conduction bands so it confirms the carrier density results in Fig. 3d.

The experimental $\sigma_{xy}^A$ is obtained by using the equation $\sigma_{xy}^A = \rho_{xy}^A/(\rho_{xx}^2 + \rho_{xy}^{A\,2})$ in Fig. 5e. The experimental $\sigma_{xy}^A$ with all gate voltages are far from $e^2/h$. Since the ratio of (Mn, Bi, Te) in our MnBi$_2$Te$_4$ is not perfect (1, 2, 4), it might be caused by the defects of the thin films, magnetic disorders from impurities, or the effect by the substrate. However, a peak of about 10 S/cm appears at $V_g = 30$ V for ~7 SL film. Because the *n-p* transition also happens around that gate voltage, we can deduce that the peak position corresponds to the bandgap. In ~3 SL, $\sigma_{xy}^A$ is even one order of magnitude smaller than that in ~7 SL. Moreover, $\sigma_{xy}^A$ changes its sign around $V_g = 30$ V. According to the theoretical $\sigma_{xy}^A$ for 3 SL in Fig. 5d, there is also a sign transition of $\sigma_{xy}^A$ around $E_F = 0.12$ eV. The reason for the sign transition is in the following. For all three calculated films (Figs. 5a-c), the first two conduction bands are dominated by spin minority or spin-down component that corresponds to the two surface states (ss) of the films. The spin-down dominated conduction bands lead to the negative-sign AHE. Around 0.1~0.15 eV above the Fermi energy, one band for 3 SL, three bands for 5 SL, and five bands for 7 SL correspond to bands of inner-layers and they are dominated by spin majority or spin up component. The change of the spin component at the Fermi energy (see spin polarization in Fig. S15) leads to the sign change of $\sigma_{xy}$. Therefore, $V_g = 30$ V of ~3 SL film is mapped to $E_F = 0.12$ eV of 3 SL (Fig. 5a). The mapping from the gate voltage of ~3 SL and ~7 SL to the Fermi energy of 3 SL and 7 SL are labeled, respectively, as shown in Figs. 5a and c. While in ~3 SL film, gating leads to all *n*-type conduction, gating in ~7SL sample, however, crosses the bandgap and causes the *n-p* transition.

## Conclusion

In summary, by precisely controlling the flux ratios, wafer-scale high-crystalline MnBi$_2$Te$_4$ films have been directly deposited on Al$_2$O$_3$ and SrTiO$_3$ substrates by the co-evaporation process. We demonstrate the antiferromagnetism, perpendicular magnetic anisotropy, and anomalous Hall insulator state via temperature- and thickness-dependent AHE analysis. Our further study on the gating control of SrTiO$_3$-substrated samples identifies the *n-p* transition and ambipolar transport in ~7 SL films and the sign-reversal behavior of AHE in ~3 SL films. Both display quadratic relations of $R_{xy}^{AS}$ and $R_{xx}^0$, indicating the intrinsic Berry-curvature-induced AHE. The band structure and Berry curvature calculations show that the anomalous Hall conductivities qualitatively match the experimental values. We also attribute the sign reversal of AHE to the competing Berry curvature between polarized spin-minority-dominated surface states and spin-majority-dominated inner-bands. In addition, the mapping from the gate voltages to the Fermi levels of the band structures is provided. Our study provides the synthesis of complex materials and presents an approach to control AHE, which may facilitate the exploration of new physical mechanisms.

## Methods

### Thin-film synthesis and characterization

MnBi$_2$Te$_4$ films were grown on (001) Al$_2$O$_3$ and (111) SrTiO$_3$ substrates in Perkin Elmer 430 MBE with the *in-situ* RHEED facility. Al$_2$O$_3$ and SrTiO$_3$ substrates were initially degassed at 600 °C and 500 °C for half an hour respectively and then decreased to the aimed temperature for the growth. Mn (99.99%), Bi (99.999%), and Te (99.9999%) three elements were co-evaporated from Knudsen cells with the source temperature of 686 °C, 460°C, and 287 °C, respectively. The flux of each element was calibrated by the crystal monitor. Before taking samples out of the chamber, 2 nm Al were capped for protection. The structural characteristics were studied by XRD (Bruker D8 Discover) and TEM (FEI Tecnai F20). The cross-section TEM sample was prepared using FIB (FEI Scios DualBeam). The films' thickness on SrTiO$_3$ was measured by AFM (Park NX10) as the substrates were too small to be directly monitored by *in-situ* RHEED during the growth.

### Electrical and magnetization measurement

For the magneto-transport experiment, thin films were cut and confined to the six Hall-bar structures and measured by Physical Properties Measurement System (PPMS, 9 T) and Oxford System (TeslatronPT, 12 T), with data collected by SR830 and Agilent B2912A. The magnetization measurements were performed in DC-Superconducting-Quantum-Interface-Devices (SQUID, 7 T).

### Band structure calculations

Total energies and electronic structures of MnBi$_2$Te$_4$ bulk and thin films were calculated from first-principles calculations within the framework of density functional theory using the projector augmented wave pseudopotential[45,46] as implemented in VASP[47,48]. The generalized gradient approximation of Perdew, Burke, and Ernzerhof[49] was used for the exchange-correlation energy, and the Hubbard *U* method[50] with $U = 4.0$ eV and $J = 0.9$ eV was applied on the Mn(3*d*) orbitals. An energy cutoff of 600 eV for the plane-wave expansion was used. Non-collinear magnetism calculations with spin-orbit coupling included were employed. The Γ-centered *k*-point mesh of $9 \times 9 \times 9$, $7 \times 7 \times 7$, and $9 \times 9 \times 1$ in the Brillouin zone (BZ) was adopted for the calculations of trigonal ferromagnetic bulk, trigonal antiferromagnetic bulk and thin-film structures by using a supercell slab with a 13.57Å vacuum layer along *z* direction, respectively. After we obtained the eigenstates and eigenvalues, a unitary transformation of Bloch waves was performed to construct the tight-binding Hamiltonian in a Wannier function (WF) basis by using the maximally-localized Wannier functions method[51] implemented in the Wannier90 package[52]. WF-based Hamiltonian has the same eigenvalues as those obtained by first-principles calculations from -1.0 ~ 1.0 eV to the Femi level. The intrinsic anomalous Hall conductivity for thin films was calculated using the WF-based Hamiltonian based on Berry curvature with a $640 \times 640$ k-mesh in two-dimensional BZ.

## Acknowledgments

This work was supported by the National Natural Science Foundation of China (11934005 and 11874116), the National Key Research and Development Program of China (Grant No. 2017YFA0303302 and 2018YFA0305601), the Science and Technology Commission of Shanghai (Grant No. 19511120500), the Shanghai Municipal Science and Technology Major Project (Grant No. 2019SHZDZX01), the Program of Shanghai Academic/Technology Research Leader (Grant No. 20XD1400200). E.Z. acknowledges support from China Postdoctoral Innovative Talents Support Program (Grant No. BX20190085) and China Postdoctoral Science Foundation (Grant No. 2019M661331). S.L. acknowledges support from China Postdoctoral Science Foundation (Grant No. 2020TQ0080 and 2020M681138). We acknowledge Zhengcai Xia and Jinglei Zhang for assistance in high-field experiments.

## Author contributions

F.X. conceived the ideas and supervised the overall research. S.L. and Z.L. synthetized the $MnBi_2Te_4$ films and measured the XRD. S.L., E.Z., Y.Z. L.L. M.Z., P.L., X.C. and X.K. performed the magneto-transport measurement and analyzed the data. S.L. and E.Z. carried out the SQUID measurement and analyzed the results. Q.S. and J.Z. did the TEM characterization. J.Y. and J.Z. performed the DFT calculations and the theoretical analyses. S.L., J.Y. and F.X. wrote the paper with assistance from all other authors.

## Competing financial interests

The authors declare no competing financial interests.

## Data availability

The data that support the plots within this paper and other findings of this study are available from the corresponding author upon reasonable request.

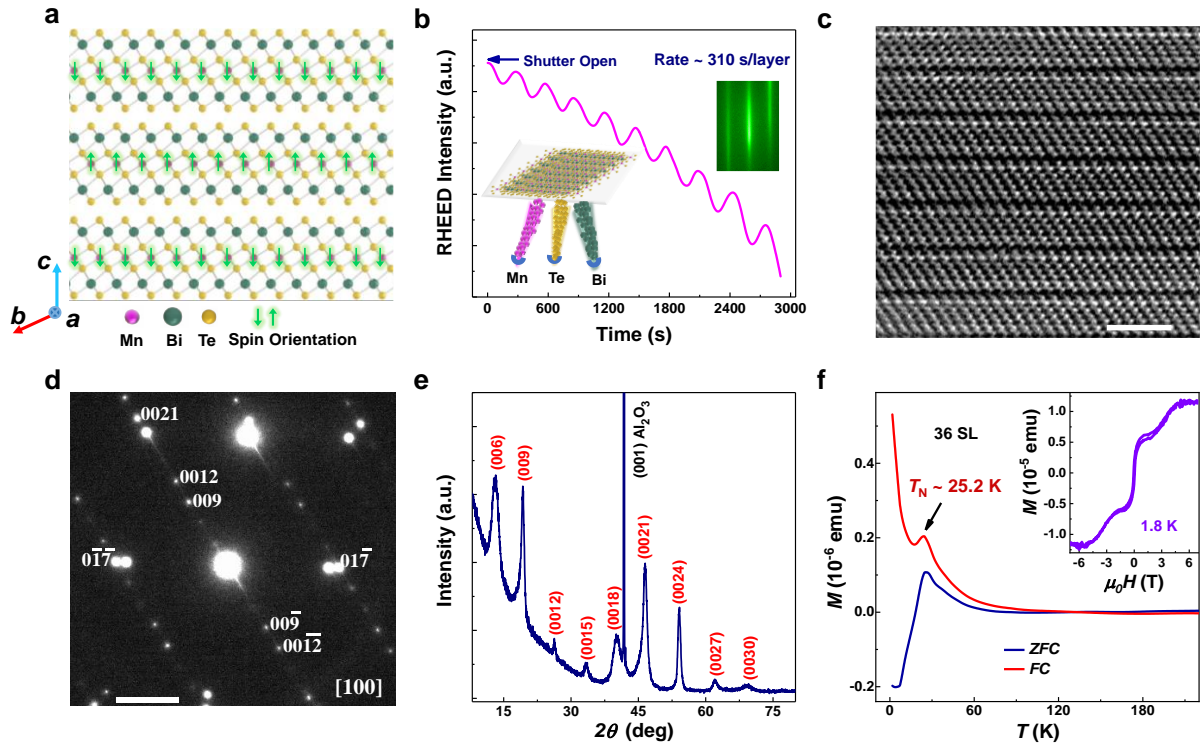

**Figure 1 | Crystal structure and characterization of a 14 SL MnBi$_2$Te$_4$ film on (001) Al$_2$O$_3$ substrate.** (a) Crystal structure of A-type AF MnBi$_2$Te$_4$ along the *a*-axis and the spins of Mn are marked with green arrows. (b) Layer-by-layer epitaxial mode observed from the RHEED oscillations. The left inset is a schematic illustration of the co-evaporation growth process of MnBi$_2$Te$_4$ films. The sharp RHHED pattern in the right inset suggests a smooth surface. (c) Cross-section HRTEM image along the *a*-axis. The scale bar is 2 nm. (d) Corresponding SAED pattern taken from [100] zone axis. The scale bar is 2 nm$^{-1}$. (e) The XRD spectrum, series of <003> peak showing the epitaxial orientation along *c*-axis. (f) *ZFC-FC* curves of magnetization as a function of temperature. Inset, Magnetization hysteresis of 36 SL MnBi$_2$Te$_4$ film at 1.8 K.

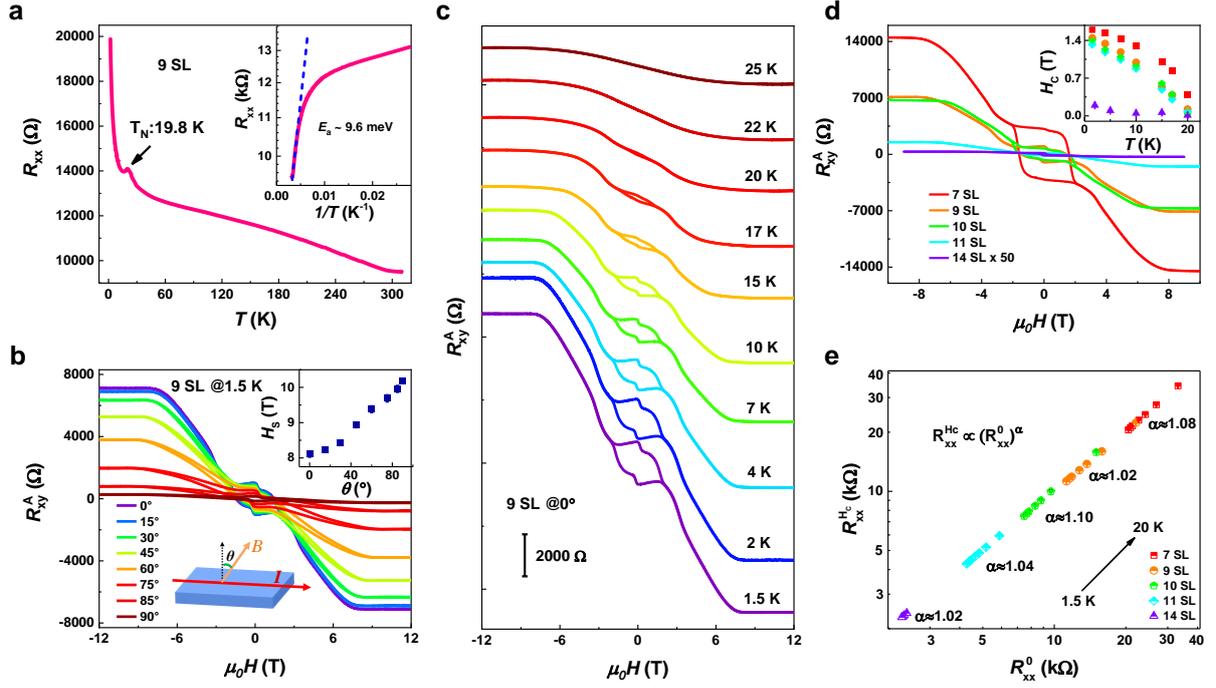

**Figure 2 | Anomalous Hall insulator state in MnBi$_2$Te$_4$ films on Al$_2$O$_3$ substrate.** (a) Temperature-dependence resistance $R_{xx}$. The kink point at 19.8 K corresponds to the magnetic transition temperature $T_N$. Inset is the fit to $R_{xx} \sim \exp(E_a/k_B T)$ with the excitation energy $E_a = 9.6$ meV. (b) Angle-dependent AHE curves. Schematic is the direction of the magnetic field with angle $\theta$. Inset is the saturation magnetic field ($H_s$). (c) AHE curves versus temperature with $\theta = 0°$. Inset shows the corresponding $R_{xx}$ at 1.5 K. (d) Thickness-dependent AHE. With the thicknesses decreasing, the hysteresis becomes broader. Note that the lowest temperature of 14 SL film was 2 K and others were measured at 1.5 K. (e) Evolution of $R_{xx}^{H_c}$ versus $R_{xx}^{0}$ at different temperatures for 7 SL, 9 SL, 10 SL, 11 SL, 14 SL MnBi$_2$Te$_4$ samples.

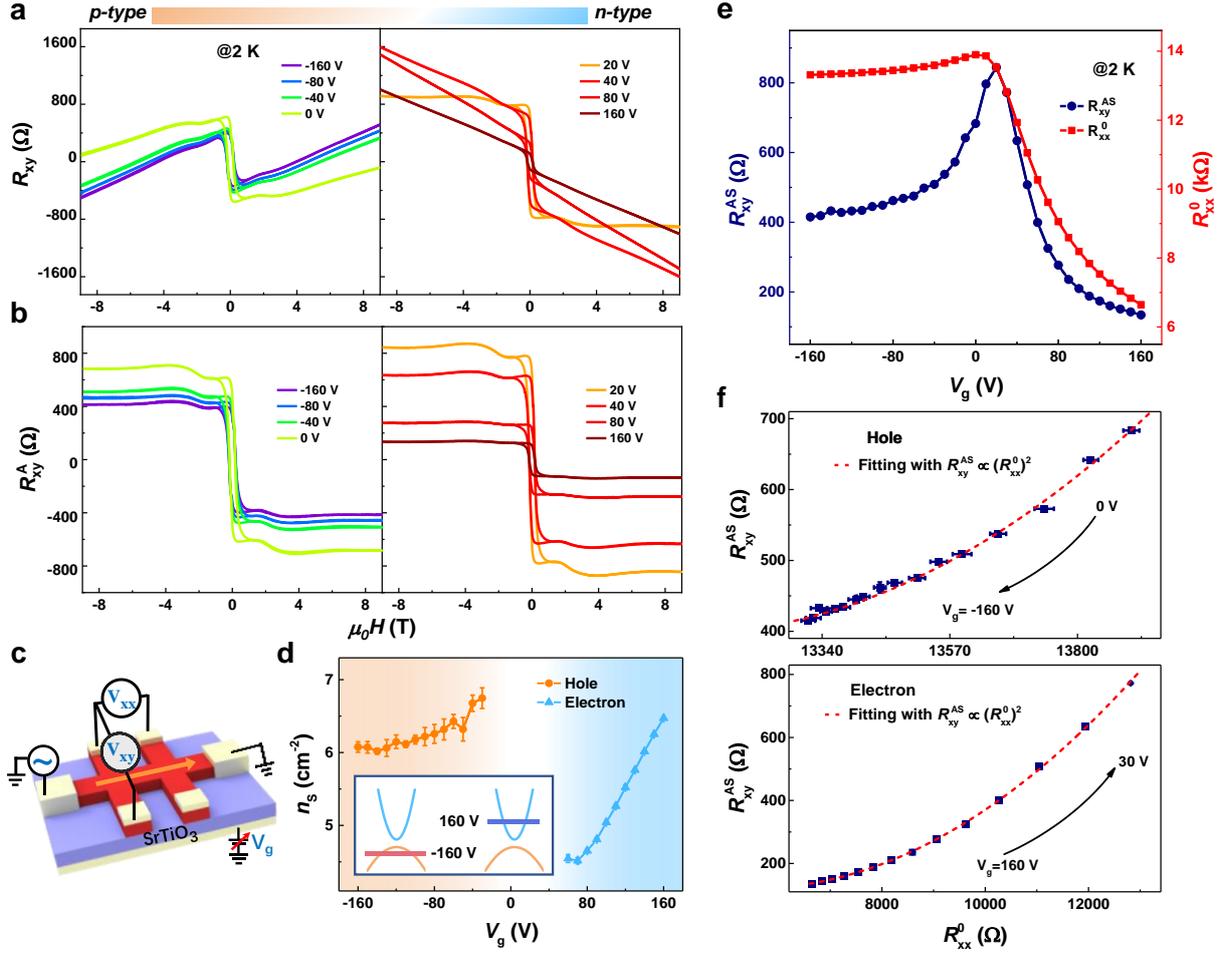

**Figure 3 | Intrinsic Berry-curvature-dominated AHE of ~7 SL MnBi$_2$Te$_4$ on SrTiO$_3$ substrate.** (a) Representative $R_{xy}$ results under different gate voltages ($V_g$), with the n-p transition occurred at ~20 V. (b) The corresponding AHE results ($R_{xy}^A$) after liner subtraction from the normal Hall effect. Negative-sign AHE curves are preserved in both the electron- and hole-dominated regions. (c) Schematic diagram of six Hall-bar geometry under the gating effect. (d) Gate-dependent sheet carrier density ($n_s$). Inset, the Fermi level tuned from conduction band (blue line) to the valence band (orange line) as $V_g$ changes from 160 V to -160 V. (e) Ambipolar $R_{xy}^{AS}$ and $R_{xx}^0$ as a function of $V_g$. (f) $R_{xy}^{AS}$ versus $R_{xx}^0$ in the hole-dominated and electron-dominated regions, the carmine dashed lines are the fits to the scaling law of $R_{xy}^{AS} \propto (R_{xx}^0)^2$. The preserved negative AHE curves in (b) and quadratic relation in (f) suggest that this AHE comes from the Berry curvature effect.

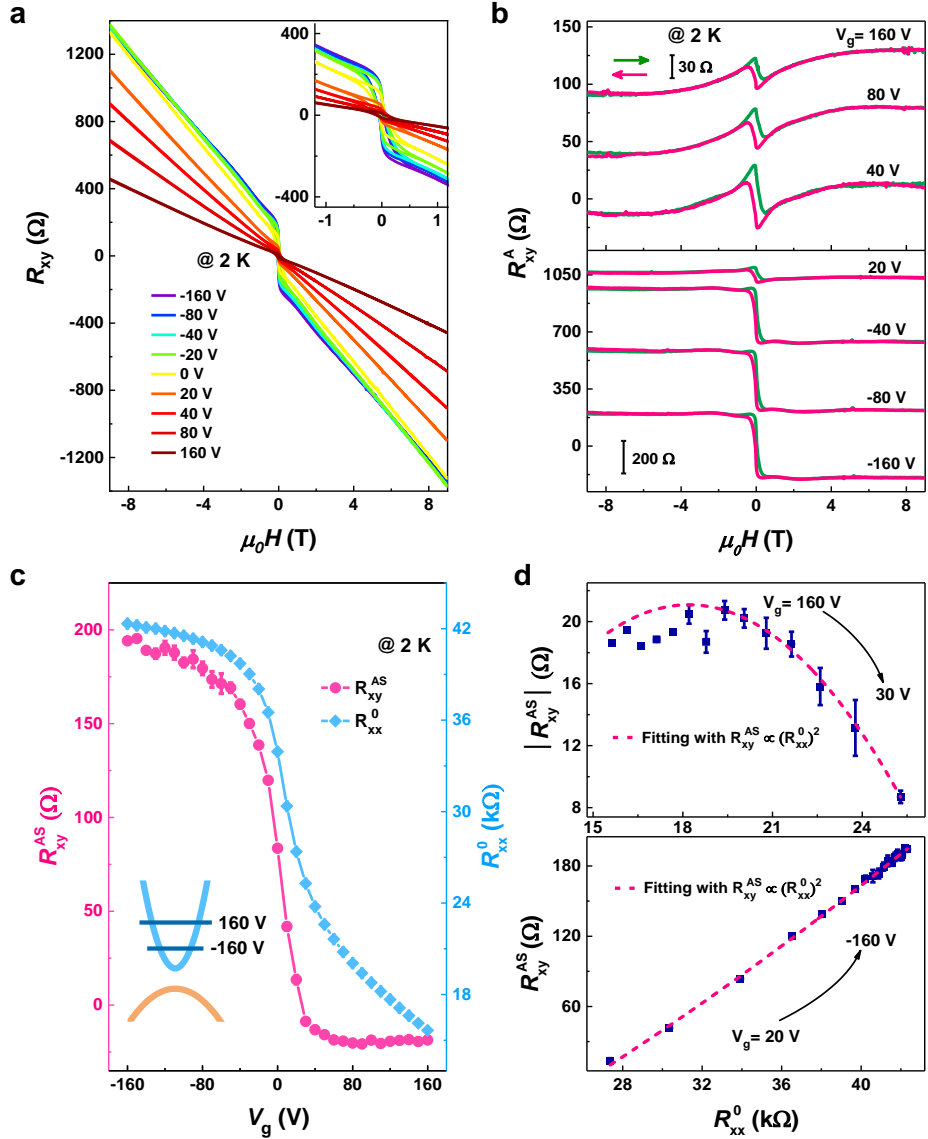

**Figure 4 | AHE sign reversal in ~3 SL MnBi$_2$Te$_4$ on SrTiO$_3$ substrate.** (a) Hall resistance under different gate voltages $V_g$, behaving *n*-type conduction in the whole gating range. Inset, the enlarged $R_{xy}$ to show the hysteresis. (b) AHE curves after linear subtraction. The transition of AHE-sign appears at $V_g$=30 V. (d) $R_{xy}^{AS}$ and $R_{xx}^0$ versus $V_g$, both showing the monotonic trend. Inset, schematic diagram for the Fermi level modulation. (d) Quadratic scaling relation between $R_{xy}^{AS}$ and $R_{xx}^0$ in the negative-sign and positive-sign AHE regions.

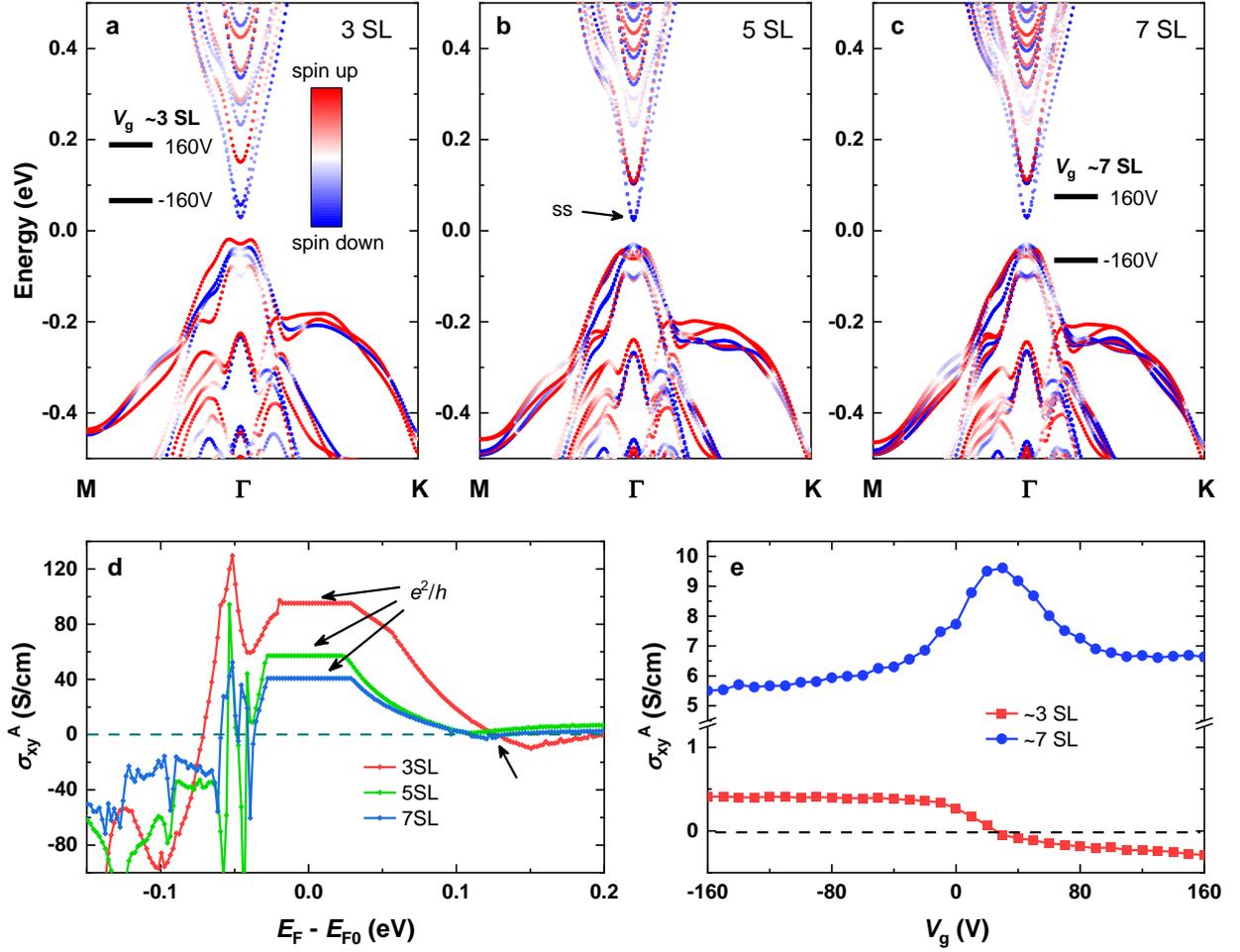

**Figure 5 | Band structures and anomalous Hall conductivities ($\sigma_{xy}^A$) for odd-layered MnBi$_2$Te$_4$.** (a)-(c) Spin-resolved band structure of 3 SL, 5 SL, and 7 SL MnBi$_2$Te$_4$. The Fermi energy is set to zero. In (b), the surface states (ss) are labeled for the first two conduction bands. In (a) and (c), the Fermi levels under various gate voltages $V_g$ were deduced from magneto-transport results in Fig. 4 and Fig. 3, respectively. (d) Calculated $\sigma_{xy}^A$ as a function of the Fermi energy. $E_{F0}$ is the calculated intrinsic Fermi energy. Note that a transition of $\sigma_{xy}^A$ from positive to negative appears around 0.12 eV for 3 SL. (e) Experimental $\sigma_{xy}^A$ for ~3 SL and ~7 SL films as a function of gate voltage $V_g$.